\documentclass{article}
\usepackage{amssymb}
\usepackage{amsmath}
\usepackage{physics}
\usepackage{graphicx}
\usepackage{subfig}
\usepackage{tikz}
\usepackage{bm}
\usepackage{authblk}

\usetikzlibrary{shapes.geometric, arrows}
\graphicspath{ {./images/} }
\tikzstyle{startstop} = [rectangle, rounded corners, minimum width=3cm, minimum height=1cm,text centered, draw=black, fill=red!30]
\tikzstyle{io} = [trapezium, trapezium left angle=70, trapezium right angle=110, minimum width=2cm, minimum height=1cm, text centered, text width=3cm, draw=black, fill=blue!30]
\tikzstyle{process} = [rectangle, minimum width=3cm, minimum height=1cm, text centered, text width=3cm, draw=black, fill=orange!30]
\tikzstyle{decision} = [diamond, minimum width=3cm, minimum height=1cm, text centered, text width=4cm, draw=black, fill=green!30]
\tikzstyle{arrow} = [thick,->,>=stealth]

\begin{document}

\title{The deep learning and statistical physics applications to the problems of  combinatorial optimization.\footnote{National Supercomputing Forum (NSCF-2019), Pereslavl-Zalessky, Program Systems Institute of the RAS, Russia.}}

\author[1]{Semyon Sinchenko}
\author[1, 2, 3]{Dmitry Bazhanov}

\affil[1]{Moscow Aviation Institute, Volokolamskoe shosse, 4, Moscow, 125993, Russian Federation}
\affil[2]{Faculty of Physics, Moscow State University, GSP-1, Leninskie Gory, Moscow, 119991, Russian Federation}
\affil[3]{Institution of Russian Academy of Sciences, Dorodnicyn Computing Centre of RAS, Vavilov St. 40, Moscow,  119333, Russian Federation}

\maketitle

\begin{abstract}
 We present herein a new approach based on the simultaneous application of the deep learning and statistical physics methods to solve the combinatorial optimization problems. The recent modern advanced techniques, such as an artificial neural network, demonstrate their efficiency for solving various physical tasks for the quantum many-body systems, which may be related directly to the problems of combinatorial optimization. One of them is a classical Maximum Cut (MaxCut) problem, which we ascribe here to the search of the ground state of a quantum many-body system using an artificial neural network and deep learning. We found that the exact solution received for a quantum system corresponds to its counterpart in a classical MaxCut problem. As a proof, we have realized our approach for two random graphs of different size, containing 60 vertices (885 edges) and 100 vertices (2475 edges), and achieved for them the total performances $p_{60} \sim 0.99$ and $p_{100} \sim 0.97$ of known maximal cuts, respectively. 
\end{abstract}

\section{Introduction}
\subsection{The MaxCut problem}
The MaxCut problem is nowadays a well-studied problem of combinatorial optimization and integer programming. The problem can be defined as follows:
for a given Graph $G = (V, E)$\footnote{Herein, we consider for simplicity the  unweighted and undirected graphs only. However, we will discuss finally about the generalization of our ansatz.}, where $V$ is the set of vertices and $E$ is the set of edges, the MaxCut problem seeks to find the subsets of vertices $V_1$ and $V_2$ in such a way that the number of edges, which incident to vertices from different sets, are maximal:
$$\sum_{(i,j) \in E} (1 - x_ix_j) \to Max,$$
where $x_i = 1$ if the vertex from $V_1$ set and $x_i = -1$ if the vertex from $V_2$ one.

In order to describe MaxCut problem one can define also the \textit{performance} parameter $p_A$ for a given algorithm $A$ that exhibits the cut of size $C$ close to the optimal value of cut $MCS$:
$$p_A = \frac{C}{MCS}$$

\subsection{The well known approaches}
\begin{figure}
\centering
\begin{tikzpicture}[thick,scale=0.8, every node/.style={scale=0.8}, node distance=6cm]
\node (sample) [startstop] {New permuted sample};
\node (energy) [io, right of=sample] {Calculate the energy};
\node (accept) [decision, below of=energy] {Accept with probability \\ $p \sim P_{Gibbs}(E_{cur}, E_{new}, T)$};
\node (temperature) [process, below of=sample] {Decrease\\ the temperature};

\draw [arrow] (sample) -- (energy);
\draw [arrow] (energy) -- (accept);
\draw [arrow] (accept) -- (temperature);
\draw [arrow] (temperature) -- (sample);
\end{tikzpicture}
\caption{The flowchart of Simulated Annealing process.} \label{fig:SA}
\end{figure}
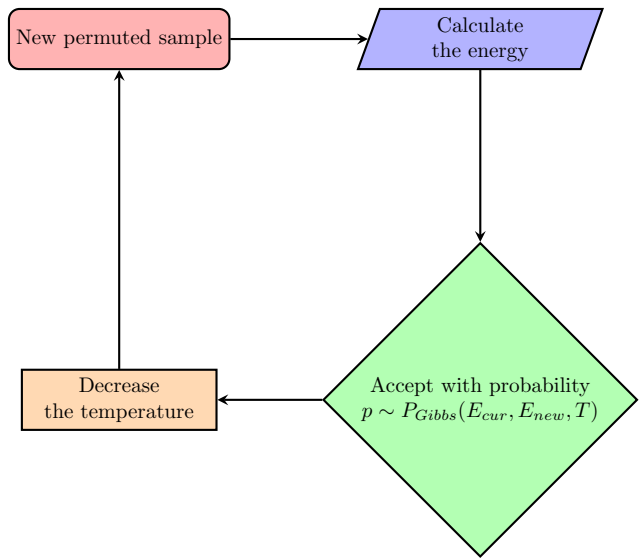

The MaxCut problem is known to be NP-hard problem and its exact solution, that can be found by the branch-and-bound algorithms, is unfeasible in the case of medium or large graph. Some of these algorithms are successful in solving the MaxCut problem, but they achieve the worth solution for performance parameter about $ p_A \sim 0.90$ \cite{MaxCutClassic}. Besides that, there was a theoretically predicted upper bound that guarantees the performance of algorithms to be not more than $p_{A}^{b} \sim 0.941$\cite{approxBound}.
Nevertheless, during the last years the novel approaches based  on the deep Neural Network (NN) and Machine Learning (ML) have been developed and intensively applied for solving the combinatorial optimization problems\cite{Bengio}. Some of them \cite{GCNSolver}\cite{PDP} were employing, so-called,  Graph Convolutional Networks (GCN), which represent the new class of NN introduced specifically to solve the graph semi-supervised learning problems\cite{GCN}. Another ones, based on Reinforcement-Learning (RL)\cite{RLopt}, were involving  the action space and reward for the applied algorithm. The latter helps the algorithm to learn and seek the best actions that maximize the reward. Besides, there is a wide class of algorithms, which are built up on Monte Carlo Markov Chains (MCMC) \cite{MCMC}. This class of algorithms does not guarantee anyone the nice solution in a reasonable amount of time, but demonstrates a very good performance in some combinatorial optimization problems, when the built-in hyperparameters are properly tuned. One of the best algorithm of this class is a well known and named as Simulated Annealing (SA)\cite{simAnn}. In the framework of SA one can employ the random permutation of the current state at each iteration step. In our case this state can be ascribed to $V_1$ or $V_2$ set. Then a new state is drawn from the probability distribution $p$:
$$p = P_{Gibbs}(E_{cur}, E_{new}, T),$$ 
where $E_{cur} (E_{new})$ presents the current (new) state endowed by some energy variable and  $P_{Gibbs}$ is a conventional Gibbs distribution, written as:

$$P_{Gibbs}(E_{cur}, E_{new}, T) = \begin{cases}
	1,& \mbox{if } E_{new} \geq E_{cur}\\
	\exp{\frac{E_{new} - E_{cur}}{T}}, &\mbox{otherwise}
\end{cases}$$

where $T$ is a temperature parameter, which should be tending slowly down, for example as a function ($T \sim \frac{1}{N}$) of learning steps $N$. For the MaxCut problem the energy variable in SA algorithm can be defined as a cut size of the considered graph. A flowchart in the  Fig.\ref{fig:SA} illustrates a workflow of SA process.

However, there were some cases reported before where the treatment based on SA algorithm failed to find a good solution in a reasonable amount of time \cite{simAnnFail}. Specifically, some difficulties can arise within MCMC-based algorithms due to their poor convergence at meaningful temperatures for real conditions. Therefore the main goal of this work is an attempt to eliminate such drawbacks of MCMC-based algorithms and to improve them by implementation of the latest achievements in statistical physics and quantum theory regarded to the problems of combinatorial optimization.

\section{Brief description of our ansatz}
In this work we present a novel physical-inspired approach to solve the MaxCut problem for graphs by means of deep NNs. In contrast to GCN-based methods, this approach can be easily generalized to a wide class of optimization problems and not only for graphs, but for any other instances. In comparison to RL-based methods we suggest to use more simple NN with the classical gradient descent and minimization procedures that leads to more stable learning process than RL-learning process. 
In our approach we treat the problem of combinatorial optimization as a physical task that should be solved for quantum many-body system with a special, so-called, {\em cost} Hamiltonian by means of Deep Learning based techniques.

\subsection{Quantum many-body problem}
The searching of the wave function $\Psi$ of the many-body system is one of the most important, but one of the hardest problems in quantum physics. The wave function contains the whole information about the state of any quantum many-body system (particles, spins, etc.) and the dynamic of the wave function allows to predict the state of such system at the each moment. Formally the wave function $\Psi$ is the complex-valued function and its square modulus relates to the probability to find the system at state $\ket{\psi}$ (here $\ket{x}$ is a ket notation for the vector in a complex Hilbert space) as the result of the measurements. The wave function of the system can be found analytically as a solution of the Schr\"{o}dinger equation: $$i\hbar\frac{d}{dt}\ket{\Psi(t)} = \hat{H}\ket{\Psi(t)},$$
where $\hat{H}$ is the Hamiltonian operator that involves all interactions between particles of the whole system. Therefore the amount of information that is necessary to be stored for full encoding of  the system increases exponentially with the number of particles. Thereby, the necessity in computation the integrals or sums (in discrete cases) over all Hilbert space fails to use the direct approaches for solving the problems of quantum many-body systems. One of the most popular solution is a Quantum Monte Carlo method that uses the variations of the Monte-Carlo method to estimate integrals over configuration space in polynomial time\cite{VMC}. Another popular approach named Matrix Product State is the generalization of numeric renormalization-group procedure to quantum lattice, and also the more general Tensor Network approach. However, there are some instances of quantum systems where the mentioned approaches fail\cite{VMCfail}.

\subsection{The link between  MaxCut and quantum Hamiltonian}
Several attempts to solve the quantum many-body problem by means of combinatorial optimization methods are known from the state-of-the-art study\cite{MaxCutHam}. 

Let us assume that any arbitrary quantum system, that consists of $|V|$ particles with quantum spins $s = \{+1, -1\}$, is ascribed for a given graph $G$ and defines the cost Hamiltonian for this graph in the following manner:
$$\hat{H} = \sum_{(i,j) \in E}A_{ij}\sigma^z_i\sigma^z_j,$$
where $A$ is the adjacency matrix of the graph and $\sigma^z$ is the Pauli matrix:
$$\sigma^z  = \begin{pmatrix}
	1& 0\\
	0& -1
\end{pmatrix}$$
The Hamiltonian is diagonalizable in $z$-basis that allows to  solve the MaxCut problem more efficiently.

It is well known, that the ground state of a quantum system is the state with minimum energy (or minimized Hamiltonian). Thereby, if from the given graph the subsets of $V_1$ and $V_2$ vertices are determined as $V_1 = \{x_i: s_i = 1\}$ and $V_2 = \{x_i: s_i = -1\}$, where $s$ is the particle spin, then the search for the ground state energy $\Large\varepsilon_{0}$ of the quantum system with Hamiltonian $\hat{H}$ leads to the solution of the MaxCut problem:

$$C = -\frac{\Large\varepsilon - \sum_{(i,j) \in E} 1}{2},$$

where $\Large\varepsilon$ is an eigenvalue of $\hat{H}$. 
The reverse statement is also true: if the MaxCut problem is solved for the given graph, the ground state of the corresponding quantum system is also found. Such approach is being actively developed nowadays in the area of Quantum Information Theory and Quantum Algorithms and shows an excellent results in experimental quantum computations for small graphs\cite{QAOA}\cite{QA}.However, no really scalable industrial quantum computer still exists.

\subsection{Neural Quantum States}
The new approach to solve the quantum many-body problem, which was  reported in \cite{CarleoScience}, is based on the Neural Quantum States (NQS). According to this work, one can define a NN with one hidden layer as function $F(\bf{S})$ where $S$ is the spin configuration vector: $S = \{s_i\}$. That function is fully described by the matrix of weights $\bf{W}$, vector of biases $\bf{b}$ and the activation functions $\sigma$\footnote{For example sigmoid-function: $\sigma(x) = \frac{1}{1 + e^{-x}}$}:
$$F(\bf{S}) = \sigma(\bf{b} + \bf{W}\vdot{\bf{S}})$$
Owing to Cybenko universal approximation theorem any function $[0, 1]^n\to{[0, 1]}$ can be approximated by NN with one hidden layer with any precision\cite{Cybenko}. There is the generalization of that theorem to complex plane\cite{ComplexCybenko}. 
Thus, hypothetically, for every real quantum system, there is a NN with weights $\bf{W}$ and biases $\bf{b}$, so that the output of this network approaches the $\Psi$ function of this system.

We can define the deep NN (in our case it was a simple Feed-Forward Neural Network (FFNN)), which brings us a {\em predictor}\footnote{A black-box model that receives a state as input and returns a $\Psi$ function as output}. This predictor can give the probability of any state of our system. Thereby, we can get the samples of system configurations $S=\{s_{i}\}$ drawn from the probability distribution defined by our variational $\Psi(S)$ function by means of this predictor and employing MCMC-methods\cite{MCMC}.
From physical principles, we know that every physical system strives to have a gain in energy and tends to its ground state with minimum energy. It means that we will get the states close to energy minimum if we draw the samples of system configurations from the true $\Psi$ function. Consequently, this allows us to determine the energy-minimization procedure:
\begin{itemize}
\item To draw the samples of system configurations $S=\{s_{i}\}$ from the probability distribution by means of predictor and employing FFNN; 
\item To estimate an energy of the samples $E_{loc}(S)$ and the stochastic expectation value of this energy $\langle \hat{H} \rangle \simeq \langle\langle{E_{loc}}\rangle\rangle$;
\item To estimate the gradient of energy $\partial_{p} \langle{\hat{ H}}\rangle$ (see in Appendix $A$ for more details);
\item To update the weights $\bf{W}$ and biases $\bf{b}$ of FFNN by using the Gradient-Descent methods;
\end{itemize}

\subsection{In comparison with Simulated Annealing}

In principle, the closest approach to our NQS-based ansatz is SA algorithm, where the heuristic Gibbs distribution is essentially replaced
 by the FFNN, which receives a state as input and returns a complex-valued $\Psi$ function as output. We use the squared modulus of $\Psi$ function to calculate the acceptance probability $p$: 

$$p = \Bigl|\frac{\Psi(S')}{\Psi(S)}\Bigr|^2$$

Using this probability, we can redefine the sampling procedure. To do so far, we apply an analogue of the annealing process (here we used Metropolis-Hastings algorithm) at each learning epoch under  implementation of deep NN and calculate the acceptance probability instead of using the Gibbs one. Besides, at the end of learning epoch we update the weights of deep NN based on stochastic reconfiguration procedure \cite{SRmethods}:

\begin{center}
   \begin{align*}
        & G = \frac{\partial_{W}{\Psi(S)}}{\Psi(S)} \\
        & E_{loc} = \sum_{S`}H_{SS`}\frac{\Psi(S`)}{\Psi(S)} \text{\footnotemark}\\
        & SR = \langle{G^{T}G}\rangle - \langle{G}\rangle \langle{G}\rangle^T \\
        & F = \langle{G E_{loc}}\rangle - \langle {G} \rangle \langle {E_{loc}}\rangle
        \\
        & W_{n+1}=W_{n}-\alpha SR^{-1}F
        \text{\footnotemark}
    \end{align*}
\end{center}

\addtocounter{footnote}{-2}

\stepcounter{footnote}\footnotetext{In the present case the Hamiltonian can be  diagonalized in $z$-basis: $H_{SS`} = \delta_{SS`}\hat{H}S$ and $E_{loc}(S) = \hat{H}S$.}

\stepcounter{footnote}\footnotetext{The update procedure may be more complex, for instance, with Nesterov momentum}

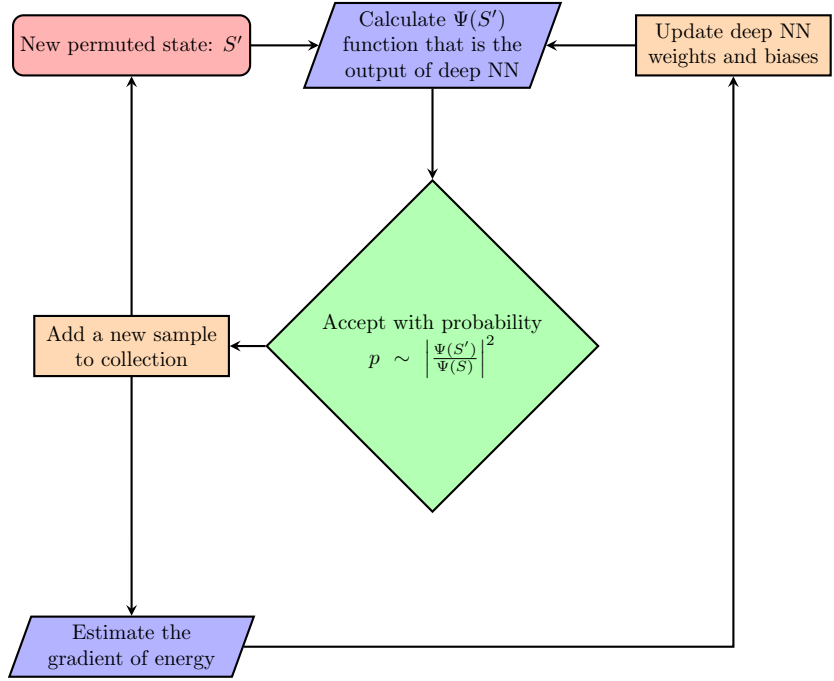
\begin{figure}
\centering
\begin{tikzpicture}[thick,scale=0.8, every node/.style={scale=0.8}, node distance=5cm]
\node (sample) [startstop] {New permuted state: $S'$};
\node (nn) [io, right of=sample] {Calculate $\Psi(S')$ function that is the output of deep NN};
\node (accept) [decision, below of=nn] {Accept with probability \\ $p \sim \Bigl|\frac{\Psi(S')}{\Psi(S)}\Bigr|^2$};
\node (collect) [process, below of=sample] {Add a new sample to collection};
\node (gradient) [io, below of=collect] {Estimate the gradient of energy};
\node (update) [process, right of=nn] {Update deep NN weights and biases};

\draw [arrow] (sample) -- (nn);
\draw [arrow] (nn) -- (accept);
\draw [arrow] (accept) -- (collect);
\draw [arrow] (collect) -- (sample);
\draw [arrow] (collect) -- (gradient);
\draw [arrow] (gradient) -| (update);
\draw [arrow] (update) -- (nn);
\end{tikzpicture}
\caption{The flowchart of the developed approach.} \label{fig:QSA}
\end{figure}

We also discard some first samples of system configurations at each learning epoch to make the learning process more stable. A detailed flowchart of the approach presented is shown in the Fig.\ref{fig:QSA}.

\section{Workflow details}
\begin{figure}
\centering
\subfloat[60-vertex graph]{{\includegraphics[scale=0.2]{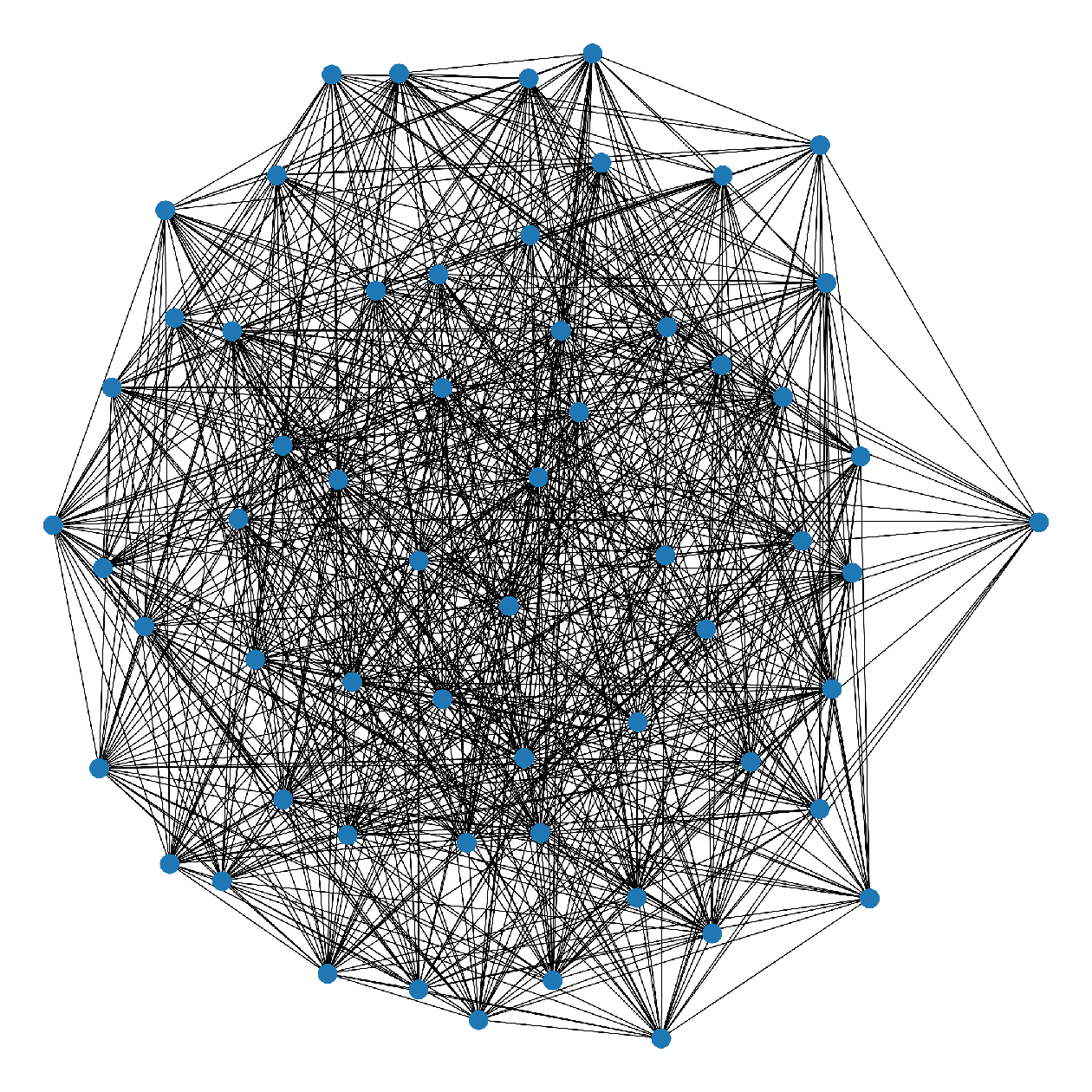}}}
\qquad
\subfloat[100-vertex graph]{{\includegraphics[scale=0.2]{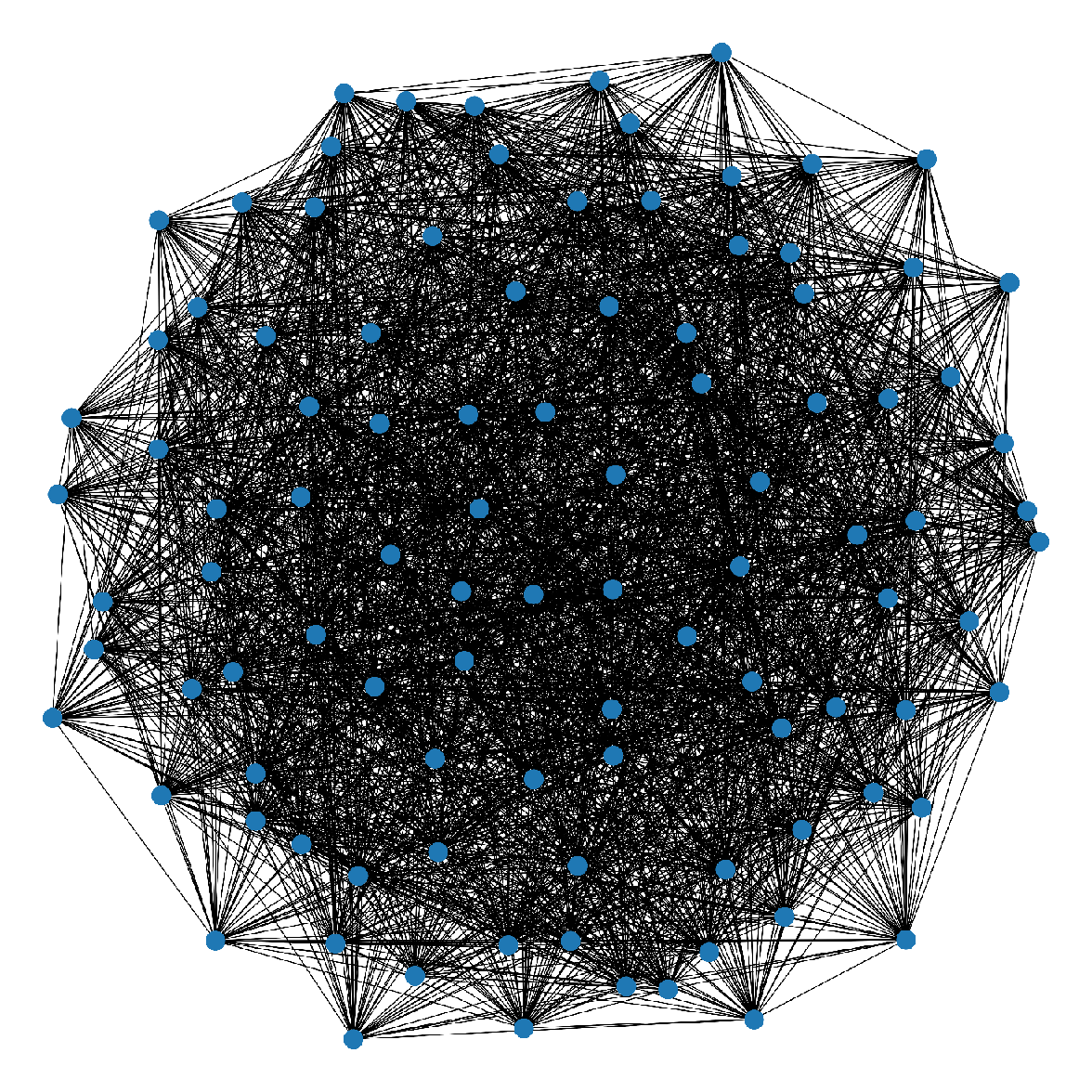}}}
\caption{The sketches of employed graphs} \label{fig::graphs}
\end{figure}

We have employed two graphs from the Biq-Mac collection: the first graph named \textbf{g05\_60.0} (60 vertices, 884 edges) and the second graph named \textbf{g05\_100.0} (100 vertices, 2474 edges)\cite{BiqMac}. Both graphs are illustrated in the Fig.\ref{fig::graphs} using NetworkX library\cite{networkx}.
For each graph the corresponding quantum Hamiltonian has been determined and the NQS approach was applied to find its ground state energy $E_0$. The detailed description of the hyperparameters for NQS is performed in Appendix $A$. 
At the end of the learning process, when deep NN yields a nearly true $\Psi$ function, we have drawn 2000 samples from the probability distribution $p$. Then the cut size ($C$) was computed for each sample and compared with benchmark solution ($MCS$) of the MaxCut for a given graph. 

We have used also the NetKet toolkit in order to estimate the ground state energy $E_0$ and $\Psi$ function of a given quantum Hamiltonian\cite{netket}. This toolkit presents a high-performance library written in $C++$ and containing the Python-bindings. The fully developed source code and Dockerfile with NetKet are available in the GitHub repository\cite{MyGitHub}.

\section{Results}

\begin{figure}
\centering
\includegraphics[scale=0.5]{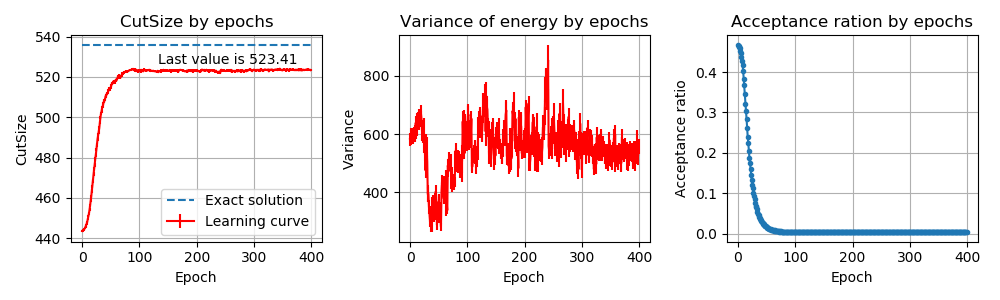}
\caption{The learning curve (left), energy variation (middle) and acceptance rate (right) for 60-vertex graph.} \label{fig:lc60}
\end{figure}

\begin{figure}
\centering
\includegraphics[scale=0.5]{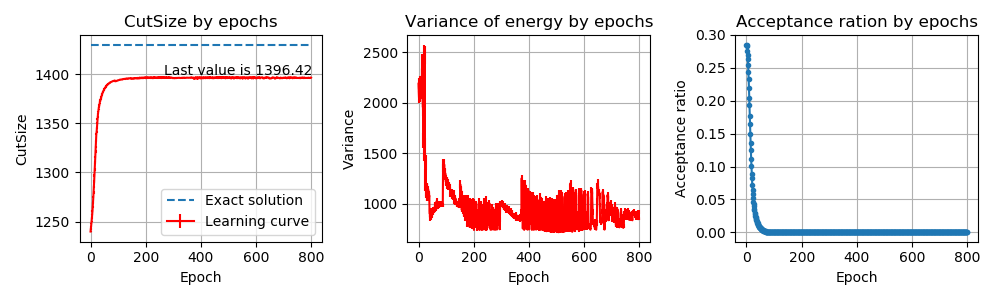}
\caption{The learning curve (left), energy variation (middle) and acceptance rate (right) for 100-vertex graph.}  \label{fig:lc100}
\end{figure}

We present in the Fig.\ref{fig:lc60} and Fig.\ref{fig:lc100} the progression curves corresponding to the learning processes that 60-vertex and 100-vertex graphs possessed, respectively.   
The left plots of both figures demonstrate the progression of cut sizes ($C$) of both graph instances, that corresponds to the expectation values of energy $\langle \hat{H} \rangle$ of the cost Hamiltonians. Moreover, the dashed lines on these plots represent the estimated  optimal solutions of MaxCut ($MCS$) for considering graphs (or exact ground state energies of the cost Hamiltonians). It is well seen in both plots that, when the learning process progresses with subsequent NN updates, the probability of sampling the high quality solutions increases rapidly for both graph instances (learning epochs $\gtrsim$ 100). The detailed energy variations with learning iterations are demonstrated in the middle plots of the Fig.\ref{fig:lc60} and Fig.\ref{fig:lc100}. It is worth to note here, that for the physical systems described by "good" Ising or Heisenberg models of Hamiltonians the energy variations should be vanished, since the samples produced during the NN learning are close to the optimal solution for the ground state with a high probability\cite{CarleoScience}. However, for a given graph the sample that differ from the ground state by only one vertex can exhibit the significant energy variations, since the degree of this vertex of a graph is not the same.  
Nevertheless, as we can see, there is no difficulty with a large energy variations, because the expectation value of the energy reaches  an optimal solution anyway. 
The right plots of the Fig.\ref{fig:lc60} and Fig.\ref{fig:lc100} depict the acceptance ratio of Metropolis-Hastings algorithm. It can be 
seen that the diversity of states decreases rapidly with the learning process of our NN. This result indicates also that we are going towards to a nearly true $\Psi$ function.

Note that in contrast to a classical deep learning, where the learning process can be controlled by an error parameter only (for instance, the energy difference between the well known value and the value estimated for the samples), in present approach there are three significantly different key parameters: error, energy variation, and acceptance rate.
If the energy expectation value is within a plateau of the learning curve and the acceptance rate is close to zero, one can conclude that we have achieved a very good fit to the true $\Psi$ function and can start the sampling from it to generate the optimal or nearly optimal solution to a given MaxCut problem.
This is a very important advantage of the approach realized in current work, since the control of the fitting process in unsupervised learning (the optimum value of cut ($MCS$) is unknown in reality) is a hard problem.
Therefore, having looked at the learning curves in the Fig.\ref{fig:lc60} and Fig.\ref{fig:lc100}, we can conclude that the learning progression up to 800 epochs is our overhead cost and can be interrupted significantly earlier after $\sim 100-200$ learning epochs.
 
\begin{figure}
\centering
\subfloat[60graph][60-vertex graph] {{\includegraphics[scale=0.45]{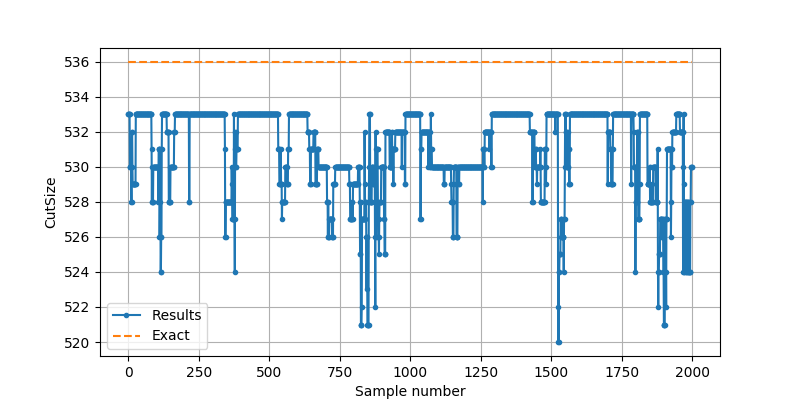}}}
\qquad
\subfloat[100graph][100-vertex graph] {{\includegraphics[scale=0.45]{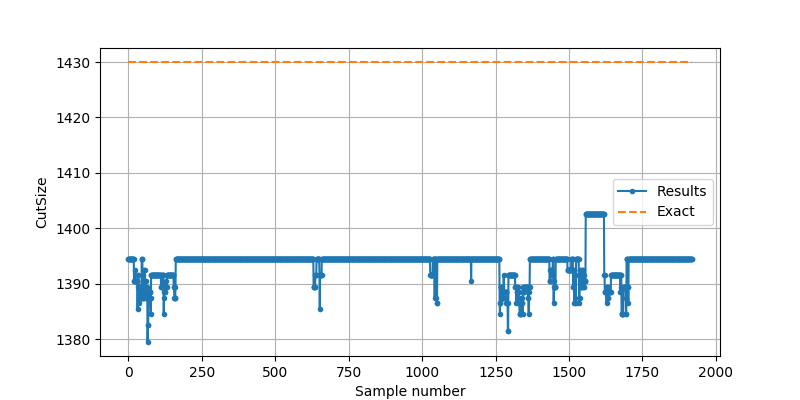}}}
\caption{The computed cut size ($C$) for 2000 samples drawn from the probability distribution of 60-vertex graph(a) and 100-vertex graph(b), respectively. The dotted line corresponds to the optimal value of cut ($MCS$) for a given graph.} \label{fig:samples}
\end{figure}

As previously claimed, we should return for solving a classical MaxCut problem in order to receive the optimal solution for a given graph based on the learning of NQS $\Psi(S)$ function. To do so, for each graph up to 2000 samples were drawn from the probability distribution corresponding to $\Psi(S)$ function using the Metropolis-Hasting algorithm. Then, these samples were used to find an optimal ordering of the high-quality solutions through the calculation of the cut size (see Fig.\ref{fig:samples}). As shown in the Fig.\ref{fig:samples}, the sampling up to 2000 samples reveals that most of the obtained solutions are the high-quality solutions, especially for 60-vertex graph. These findings result on the remarkable values of performance parameter obtained for both given graphs across the considered sets of their samplings:

$$p_{60} \sim 0.99 \quad \mbox{for 60-vertex graph}$$
$$p_{100} \sim 0.97 \quad \mbox{for 100-vertex graph}$$

We want to emphasize that the similar results one can achieve by using SA algorithm (or an other MCMC-based algorithms), but such approach can fail for more sophisticated problems than those we have studied here\cite{simAnnFail}. 
We expect that the employing of the NNs instead of Gibbs-sampling or any other prior distribution can improve the convergence significantly.

\section{Outlook}
To conclude, our findings unveil a new insight into the application 
of a novel physical-inspired NQS-based approach to solve the various combinatorial optimization problems by implementation of an artificial NN and deep learning means. The presented approach reveals a nice convergence towards a high-quality solutions of the MaxCut problem with a remarkable performance $p_A$ on two typical in size graph instances (60-vertex and 100-vertex graphs). It demonstrates a noticeable supremacy over a classical heuristic algorithms of optimization in the performance, acceptance rate, solution time, and computational efficiency of simulation procedure. Moreover, the method of NQS has а reliable theoretical background  based on quantum and perturbation theory\cite{PertTh} and has applied successfully to study the problems of the quantum many-body systems  
\cite{NQS1,NQS2,Ising1,NQS3}. In spite of that, the rigorous theoretical proof is still needed to ensure that our approach guarantees the performance of the obtained results for more complex instances or optimization problems. 

Finally, it is easy to show how to generalize our approach to the directed weighted graphs:
$$\hat{H} = \sum_{(i,j) \in E}A_{ij}\sigma^z_i\sigma^z_j$$
Herein, we can simply define $A$ as an adjacency matrix with weights taken as values and $E$ as a set of directed edges.
Thereby, in a more general sense, we can ascribe any unconstrained binary optimization problem to the problem of minimizing the cost Hamiltonian of the corresponding Ising-like spin glass model.
This kind of approach was demonstrated in some works on the application of quantum annealing and quantum optimization methods to a wide range of combinatorial optimization problems\cite{QAOAbook}.

It is worth to note, that there are also many advanced tools in the fast-growing field of deep learning that can be successfully applied in NQS-based approaches of combinatorial optimization.
For example, one can use convolutional layers\cite{Conv} instead of fully connected layers, which can significantly reduce the number of learning parameters and, therefore, reduce the Vapnik-Chervonenkis dimension of NN\cite{VC} and significantly simplify the computation complexity of learning. 
We assume that one can move from complex weights of NN to real weights, which allow us to use modern deep learning libraries, such as Tensorflow\cite{TF} and PyTorch\cite{PyTorch}. These libraries make it easy to determine the structure of NNs and to use fast GPU computing\cite{TF2}.

\section{Acknowledgments}
The authors are grateful for the financial support from Russian Foundation for Basic Research (Grant No.19-29-03051-mk).

\section*{Appendix A. Hyperparameters of NQS}

In the original paper of G.Carleo and M. Troyer the Neural Quantum States (NQS) approach was proposed to represent the wave function $\Psi$ of the many-body quantum system using the Restricted Boltzmann Machine (RBM)\cite{CarleoScience}. Using RBM there are some difficulties when the number of particles in a quantum system increases essentially.
For example, if the density of the hidden units is only four, for a graph of 100 vertices there will be 400 hidden units. Such amount of units produces 40000 weights in the hidden layer, which become an operating problem for stochastic reconfiguration method: for 10000 samples at each learning epoch the dense matrices of 40000$\cross$10000 dimension must be operated. To get around this problem we used simple Multi-Layer perceptrons instead of RBM.

\subsection*{Network Architecture for 60-vertex graph}
For small graph we used the following structure:
\begin{itemize}
\item Fully-Connected layer: input shape 60, output shape 30
\item Fully-Connected layer: input shape 30, output shape 20
\item Fully-Connected layer: input shape 20, output shape 10
\item LnCosh layer
\item Sum layer
\end{itemize}

Learning Hyperparameters:
\begin{center}
\begin{tabular}[center]{| l | r |}
\hline
Optimizer & Nesterov momentum ($lr=0.008, \beta = 0.9$) \\ \hline
Samples at each step & 6000 \\ \hline
Discarded samples at each step (first N) & 3000 \\ \hline
Learning epochs & 400 \\
\hline
\end{tabular}
\end{center}

\subsection*{Network Architecture for 100-vertex graph}
For small graph we used the following structure:
\begin{itemize}
\item Fully-Connected layer: input shape 60, output shape 40
\item Fully-Connected layer: input shape 40, output shape 30
\item Fully-Connected layer: input shape 30, output shape 10
\item LnCosh layer
\item Sum layer
\end{itemize}

Learning Hyperparameters:
\begin{center}
\begin{tabular}[center]{| l | r |}
\hline
Optimizer & Nesterov momentum ($lr=0.008, \beta = 0.9$) \\ \hline
Samples at each step & 10000 \\ \hline
Discarded samples at each step (first N) & 5000 \\ \hline
Learning epochs & 800 \\
\hline
\end{tabular}
\end{center}


\begin{thebibliography}{20}
\bibitem{TF2}
Right now we are doing tests regarding the implementation of the NQS-solver for MaxCut problems in Tensorflow 2.0, in which the abilities of the cost Hamiltonian to be diagonalized and GPU-computed are used.
\end{thebibliography}
\end{document}